# Emulating long-term synaptic dynamics with memristive devices

Shari Lim Wei, Eleni Vasilaki, Ali Khiat, Iulia Salaoru, Radu Berdan, Themistoklis Prodromakis

**The potential of memristive devices is often seeing in implementing neuromorphic architectures for achieving brain-like computation. However, the designing procedures do not allow for extended manipulation of the material, unlike CMOS technology, the properties of the memristive material should be harnessed in the context of such computation, under the view that biological synapses are memristors. Here we demonstrate that single solid-state $TiO_2$ memristors can exhibit associative plasticity phenomena observed in biological cortical synapses, and are captured by a phenomenological plasticity model called "triplet rule". This rule comprises of a spike-timing dependent plasticity regime and a "classical" hebbian associative regime, and is compatible with a large amount of electrophysiology data. Via a set of experiments with our artificial, memristive, synapses we show that, contrary to conventional uses of solid-state memory, the co-existence of field- and thermally-driven switching mechanisms that could render bipolar and/or unipolar programming modes is a salient feature for capturing long-term potentiation and depression synaptic dynamics. We further demonstrate that the non-linear accumulating nature of memristors promotes long-term potentiating or depressing memory transitions.**

In the past, artificial neural networks (ANNs) have been brain-inspired conceptions typically developed independently from neuroscience. As such, they have largely ignored biological characteristics; for instance, the key fact that synaptic connections among neurons are bounded[1], or inherently unreliable when transmitting a signal[2], having the possibility of undergoing revertible changes depending on the timing or frequency of the neuronal signals. Recent developments of dedicated hardware implementations of ANNs are leading to the design of synaptic learning mechanisms[3,4], which bare more similarities to the biological ones, compared to the methods and software algorithms driven by pure theory. Theoretical developments in the last decade further advanced the field by linking the learning theory back to the biological substrate. A key element in this direction was the use of more realistic brain cell models, spiking neurons[2,5], and novel synaptic plasticity models capturing both short- and long-term dynamics[4,6,7]. Of particular interest are cases where the design of new learning mechanisms are constrained by the limitations of hardware, when the physics of the circuits and devices used are the reminiscent of the biophysics of the biological neurons and synapses modeled[8].

Several efforts have been made to implement these mechanisms by exploiting Complementary Metal-Oxide-Semiconductor (CMOS) topologies and emerging nanoscale cells. The complexity and capability of CMOS circuits vary, with some being able to capture short-[9-11] or long-term[12] plasticity or even both[13]. And although a full CMOS approach has some benefits in terms of cost and flexibility, it typically requires considerable amount of chip area and power, thus making the integration of large-scale neural processing systems prohibitive. To alleviate challenges imposed by CMOS implementations[14], latest efforts are leveraging the attractive attributes of emerging Resistive Random Access Memory (ReRAM) cells, also known as memristors[15], in particular exploiting their simple (two terminal) architecture and small footprint[16], their capacity to store multiple bits of information per cell[17] and the low-power required for programming[18].

To this extent, memristors have been shown to be capable of qualitatively emulating long-term plasticity, such as Spike-timing Dependent Plasticity (STDP)[19] and various STDP variations[20,21], as well as short-term plasticity (STP) (let us site our work here among others). Such approaches rely on non-volatile memory-state transitions based upon phase-change[21,22] mechanisms or the diffusion of ionic-species within an active core[23-25]. To date, many groups worldwide have shown how it is possible to induce timing-dependent conductance changes in memristive devices in a way that resemble the STDP induced changes in real synapses[21-27]. However in most cases the equivalence between the physics of memristive devices and the physics governing the behavior of real synapses has been shown only at an abstract qualitative level. Our previous work focused on showing that single $TiO_2$ memristors exhibit STP-like phenomena, which can be used for spatiotemporal computation[28]. In this work we demonstrate how single $TiO_2$ memristors are also capable of capturing long term-synaptic dynamics using the same experimental protocols used to test real synapses, and reproducing the trend of a recently established data driven plasticity rule, the triplet rule[7], which exhibits a regime of spike timing dependent plasticity (STDP) behavior, where timing matters, and a regime of classical associative hebbian regime, where neurons that "fire together wire together"[29-31]. To support this, we carry out detailed quantitative comparisons, with results of electrophysiology experiments with real synapses, and with data-driven computational neuroscience models. Most noteworthy, the best-fit data are within the range of biological cortical neuronal synapses.

**Memristive dynamics**

Our memristive device qualitatively represents a synapse (inset I of Figure 1a), with its conductance corresponding to the notion of a synaptic efficacy modulated via the arrival of a spike, i.e. a pulse applied pre-synaptically to the device's top electrode (TE), shown in inset II of Figure 1a. The post-synaptic current entering the artificial neuron, from the device's bottom electrode (BE), is proportional to the memristive conductance. Figure 1a depicts a microphotograph of one of our fabricated crossbar type $TiO_2$-based memristors (fabrication details are given in Methods). The device comprises two Pt electrodes (TE and BE) that are separated by a stoichiometric $TiO_2$ active core (cross-section is shown in inset II of Figure 1a). Following an electroforming step (depicted in Figure S1), the devices' electrical characteristics were first investigated via positive/negative ±2V voltage sweeps, resulting into a bipolar mode of switching: positive sweeps cause low- (LRS) to high-resistive state (HRS) transitions, while negative ones cause HRS to LRS transitions. This biasing cycle promoted a three orders of magnitude change in conductance ($R_{OFF}/R_{ON}$) as shown in Figure 1b. Alternative modes of switching can be realized by employing larger biasing unipolar voltage sweeps, as illustrated by the results presented in Figure 1c, where the HRS to LRS transition occurs when the positive voltage sweep reaches 3V. The co-existence of bipolar and unipolar switching in $TiO_2$ ReRAM cells, also reported in the literature[32], supports the hypothesis that the switching mechanism is filamentary in nature. To test this hypothesis, we carried out a series of cyclic voltammetry measurements on pristine devices. Similar to Figures 1b and 1c, the devices are excited with a multitude of pulses of fixed duration (30ms) and interpulse time (30ms), with the amplitude of each pulse increasing each time at a fixed step (V=50mV) until switching was observed. Figure 1d illustrates the acquired OFF/ON resistive ratios for distinct voltage sweeps limits. It is shown that the switching trend (denoted as bipolar or unipolar) is contingent on the activity (pulsing events) of each device. Contrary to the bipolar mode case, where resistive switching is proclaimed via the displacement of $O^{-2}$ vacancies, this onset threshold ascertains that unipolar switching is facilitated via Joule heating[33] that either annihilates existing, or forms new, conductive percolation paths across the $TiO_2$ active core.

The corresponding current-voltage (I-V) characteristics, with the classical pinched-hysteresis memristor signature[34] is shown in Figure 2a. After electroforming, the devices are originally in a LRS and a HRS can be achieved as the sweeping voltage bias approaches a set value Vset = 1.7V. Reversing the voltage polarity, the device switches onto a LRS at $V_{reset1}$ = -1.8V (Figure 2a). This action describes completely the bipolar behavior. The pinched hysteresis curve obtained is a clear fingerprint of a memristor[34,35]. Figure 2b demonstrates three resistive states that were obtained from 50 repeated pulsing sequences, of 10µs pulse widths, as described in the corresponding figure inset. The multi-state capacity of our memristor is modeled in this case via a random circuit breaker network model[36-38], as illustrated in the corresponding inset schematics of Figure 2b to simulate the effect of local conductance changes within the active $TiO_2$ core on the overall conductance of the solid-state device. Considering that a SET potential will facilitate some local modification of the active material in the form of a conductive filament that in turn will result in a state modulation, we represent this change by altering some of the branch resistances to higher conductance values (colored lines). The number of filaments in $R_1$ and $R_2$ as well as the corresponding values for the low and high resistive branches is arbitrarily selected for matching the average measured resistive states across all cycles. The experimental and simulated results employed in the RCB model are in close correlation, validating the notion that the attained resistive states are due to filamentary formations/disruptions. Figure 2c depicts the non-linear accumulating nature of our $TiO_2$ memristors when subsequent identical voltage pulses have a decreasing effect on the modulation of the effective resistance of our prototypes. This programming method was also employed, when necessary, to set/reset our devices at intermediate resistive states throughout our experiments.

Provided that sufficient energy is delivered to the ReRAM memristor, the active core can undergo stable phase transitions, which translate into long-term changes in the conductance of the device. Figure 3a demonstrates how a presynaptic-only strong ("tetanic") stimulus of long duration can lead to LTP. This behavior is reminiscent of the presynaptic-only mediated form of LTP observed in biology[39]. In this case, the energy accumulated at the device's core, contributes to the creation of stable percolation channels. On the other hand, Figure 3b shows an example in which the same presynaptic-only stimulus leads an initially stronger synapse (higher initial conductance) to LTD, a non-volatile conductance decrease. This property of being able to induce either LTP or LTD depending on the initial value of the synaptic efficacy is commonly observed in biology, and referred to as weight normalization[40,41]. From the device physics perspective, we argue that starting the stimulation from high conductance levels (e.g., Figure 3b rather than Figure 3a) leads to accumulation of energy that saturates the available resources ($O^{-2}$ vacancies), beyond which new percolation channels are formed. In this case, supplementary energy is dissipated as Joule heating and the existing filaments are annihilated. To emphasize this behavior, the strong stimulation scheme of 9 spikes shown on Figure 3c was employed with 1μs (5μs) long pulses. It is interesting to note that the first conductance peaks of Figure 3b are in fact increasing, possibly due to the formation of some locally reduced $TiO_2$ phases, which later on however are counteracted by the annihilation of a percolation branch that results in a saturation of the response and eventually to LTD. A very similar effect was indeed reported in the mammalian neuromuscular junction, where a synaptic transmission event can be facilitating at first before being overwhelmed by depression[42].

**Associative long-term synaptic plasticity**

The long-term plasticity observed in our devices was shown to adhere with synaptic plasticity modifications produced by pair, triplet and quadruplet STDP[43] as well as by frequency dependent STDP[44] protocols. For this set of results a separate evaluating platform was designed that is shown in Figure 4. This circuit represents an analogue asynchronous implementation composed of two spike generating circuits acting as neurons and a nanoscale solid-state $TiO_2$ memristor as the synapse, denoted as M. The two pulsing circuits respectively feed voltage spikes into each end of a single element. Each spike generating circuit incorporates an mbed NXP LPC1768 microcontroller. The mBED Testbed also controls the SLAVE mBed to provide the postsynaptic spikes.

Sending a pair of pre- and post-synaptic spikes separated by a time interval T into each end of our memristor allows to simulate pair-based STDP (Figure 5a), for Δt = $t_{post}$ - $t_{pre}$ varying between ±100ms, where $t_{pre}$ and $t_{post}$ are the times that the presynaptic and postsynaptic spike signals were elicited. The applied spikes are 3.5V square pulses that are 50μs wide. The memristor's conductance was measured after each stimulating pattern of 60 spike pairs at a frequency of 1 Hz, and was used as a measure of synaptic efficacy. A positive (negative) change in conductance is observed when the pre-synaptic spike is applied before (after) the post-synaptic spike. This implies that the pre-post spike pair with Δt>0 elicits potentiation; whereas when the timing was reversed depression occurs. The percentage change in conductance ΔG was found to decrease with increasing |Δt|. Our measured results were fitted with the Voltage-Triplet rule[7,45] as if they were measurements from real synapses (details on the employed models can be found in supplementary material). Appropriately scaled measured biological synaptic data from references[44,46] are also depicted along our results to illustrate the close phenomenological resemblance with results obtained from biological synapses. We should note here that the classical STDP curve is only elicited for moderate stimulating pulses in amplitude that are above the switching threshold of the memristor, but at the same time do not trigger a unipolar mode of switching that will result only in potentiation, as illustrated in supplementary materials.

During our testing, we noted that the initial conductance of the memristor influences the resultant change in conductance after stimuli. Focusing on the pair-based STDP protocol with Δt = ±1ms, the observed synaptic modification, as indicated by a positive or negative percentage change in conductance, was recorded for a range of initial conductance values. The applied spikes are square pulses of amplitude 3.5V and pulse width 50μs. When the extent of synaptic modification was plotted against the initial conductance values, an inverse relationship can be established as shown in supplementary materials. As expected, the pre-post case (Δt = +1ms) elicited a positive percentage change in conductance resembling synaptic potentiation, while the post-pre case (Δt = –1ms) resulted in a negative percentage change in conductance equivalent to synaptic depression. For both cases, the fitted line showed that the modulus of percentage change in conductance %ΔG is inversely proportional to the initial conductance value. The use of logarithmic x-axis nevertheless indicates that the inverse relationship is non-linear. The inverse relationship between synaptic modification and initial synaptic strength of a synapse was also observed in biological synapses, with evident LTP occurring mainly in weak synapses. LTD however did not show any significant dependence on initial synaptic strength. This implies that synapses undergo potentiation until saturated, leading to negligible synaptic modification in stronger synapses[47]. It can thus be deducted that lower initial conductance will result in a more robust synaptic modification, suggesting the potential of calibrating the initial strength of the memristor to optimise the resultant conductance modification. The observed results also proposed the existence of a conductance saturation level within memristors.

To emulate triplet protocols, we extend the previously used pair pulsing patterns to accommodate three stimulating pulses. Our experimental triplet protocol consists of 60 sets of three spikes repeated at 1Hz frequency for two cases: pre-post-pre and post-pre-post, with the timings illustrated in the inset of Figure 5b. Each spike is a square pulse with amplitude 3.5V and pulse width 50μs. Experimental testing was carried out by applying the triplet protocol with different sets of spike timing intervals ($Δt_1$, $Δt_2$). Both LTP and LTD are activated in the two sets of triplet configuration and the two processes seemingly integrate in a non-linear manner. Weak depression or no change is observed when potentiation occurs first, whereas potentiation dominates substantially when it occurs after depression. This observation appears to conform with the results of Wang et al.[44] on hippocampal cultures, where *"potentiation and depression cancel when potentiation is induced first, whereas potentiation dominates when it is induced second."* We note excellent qualitative agreement of the memristive and the biological synapse, with the exception of the post-pre-post protocol for timing (15,5), (5,15). A quadruplet protocol was also employed, as in the inset of Figure 5c, comprising a repetition of 60 sets of four spikes at a frequency of 1Hz. The spikes are square pulses of 3.5V amplitude and 50μs pulse width. For positive T, a pronounced potentiation is observed when T is small. As for negative T, potentiation is induced too but the effect is comparatively smaller, especially when T is large. We can deduce that potentiation is dominant when it follows depression, similar to the observation with triplets; a remarkable agreement with the data of Wang et al.[44].

We further examined the dependence of the synaptic modification on the repetition frequency of standard spike-pairs (Figure 6). The pair of pre- and post-synaptic spikes was applied to each end of our memristor, with 60 pairs of pre- and post-synaptic spikes being repeated at regular intervals of T=1/f, where f is the frequency in Hz. The applied spikes have a 3.5V amplitude and pulse width 50μs and the complete testing involved systematically varying the frequency f with Δt fixed at ± 10 ms. The degree of potentiation is seen to increase with repetition frequency for the pre-post pairing case, with only minimal potentiation at low frequencies. On the other hand, the post-pre pairing resulted in depression at low frequencies, with a well-defined transition from potentiation (Δt =+10ms) to depression (Δt = -10ms) observed for the low-frequency range up to 30Hz, beyond which all spike pairings resulted in potentiation for all |Δt|. Our results conform to corresponding data from experiments conducted on visual cortical L5 neurons by Sjostrom et al.[46]. The minimal pre-post potentiation at low frequency range can be attributed to the spike pairs being too far apart. A large number of inputs need to be activated in synergy to produce pronounced low-frequency potentiation. As frequency increases, the spike pairs approach closer to one another; the post-synaptic spike that was precisely synchronized with a presynaptic spike within an LTD window begin to enter an LTP window for the preceding pre-synaptic spike. This could lead to the spike trains producing both potentiation and depression interactions at the same time[46]. The observation that depression ceased to exist and instead was replaced by potentiation beyond 30Hz suggests that potentiation dominates over depression when both interactions occur closely in time. From the memristors perspective, when the device is stimulated with moderate amplitude post-pre pulsing pairs at low frequencies, there is sufficient thermal relaxation in place so that the $TiO_2$ core will only endure a reversible field-driven reduction/oxidation. On the other side, as the stimuli pairs occur more sparsely, thermal effects would prevail solely imposing a reduction process and thus an overall increase in the device's conductance.

**Summary**

In this work, we presented detailed and quantitative parallels between memristive devices, biophysically realistic models of synaptic dynamics, and electrophysiology experimental results obtained from real synapses. In particular, we demonstrated that single $TiO_2$ memristors are able to exhibit the properties of both the classic STDP rule and the Hebbian rule, in agreement with experimentally observed phenomena of synaptic plasticity[46]. Our memristive device is able to reproduce quantitatively (up to a scaling factor) a number of synaptic plasticity experiments, which include, apart from the standard STDP protocol, frequency STDP protocols, as well as protocols of triplets and quadruplets[46,47]. Our artificial synapse overall exhibits properties are remarkably similar to the triplet rule[29]. Further to its ability to demonstrate long-term plasticity, we have earlier demonstrated that the same type of devices also exhibits short-term plasticity. Earlier theoretical work has attempted to explain the development of specific connectivity motifs based on the interaction of short term and long term plasticity[30,48]. We believe that a next crucial step is to show that short-term plasticity and long-term plasticity mechanism may co-exist in the same single $TiO_2$ memristor by demonstrating the formation of such motifs in a neuromorphic memristive system.

**Figure 1** Solid-state TiO$_2$ ReRAM memristors can support both bipolar and unipolar non-volatile switching for emulating long-term plasticity. A top-view of a 2x2μm$^2$ active area and 10nm thick TiO$_2$ cross-bar architecture is shown in a), with insets I and II respectively depicting cross-sections of a chemical synapse and a pristine memristor (blue denotes the Pt TE and BE that correspond to pre- and post-synaptic terminals, with green and red corresponding to Ti and O$_2$ species that can be displaced within the functional core). Bipolar and unipolar switching modalities can co-exist in such devices, as respectively captured in b) and c); with exemplar OFF/ON resistive ratios acquired after voltage cyclometry captured in d).

**Figure 2** Electrical characteristics of our memristor prototypes. Shown are: a) pinched hysteresis I-V trend that indicates a memristor signature, b) continuous cycling (200 cycles) between three resistive states with measured and simulated response according to a filamentary formation model as shown next to each corresponding state and c) demonstration of the intrinsic accumulating non-linear response of our prototypes when programmed with voltage pulses of fixed amplitude (inset).

**Figure 3** Long-term memory transitions of a single TiO$_2$ memristor. Shown are: a) long-term potentiation (LTP), b) long-term depression (LTD) and c) pulsing sequence utilized for eliciting LTP/LTD behavior.

**Figure 4** Circuit schematic of evaluating platform employed in all long-term plasticity experiments presented in this work.

**Figure 5** The memristive synapse demonstrates associative long-term plasticity in excellent agreement with biological data. Shown are: a) pair-based STDP, b) triplets protocol and c) quadruplet protocol. and d) frequency dependence of pair-based STDP. Circular markers indicate measured data of our ReRAM memristors, while triangular markers indicate scaled data from biological synapses taken from references[44]. Solid-lines and bars show the Voltage-Triplet rule fitted on the memristor measurements. Insets show the employed protocols for each case.

**Figure 6** The memristive synapse demonstrates associative long-term plasticity in excellent agreement with biological data. Shown are: a) frequency dependence of pair-based STDP. Circular markers indicate measured data of our ReRAM memristors, while triangular markers indicate scaled data from biological synapses taken from references[46]. Solid-lines and bars show the Voltage-Triplet rule fitted on the memristor measurements. Insets show the employed protocols for each case.

**Methods Summary**

All memristor prototypes exploited in this work were fabricated by the following process flow. 200nm of $SiO_2$ was thermally grown on top of 4-inch Si wafer, with 5nm Ti and 30nm Pt layers deposited via electron-gun evaporation to serve as the bottom electrodes (Ti is used as an adhesion layer). An RF magnetron sputtering system was used to deposit the active $TiO_2$ core from a stoichiometric target, with 30sccm Ar flow at a chamber pressure of $P=10^{-5}$mbar. Finally, all top Pt electrodes were deposited by electron-gun evaporation. A lift-off process was employed for patterning purposes prior each metal deposition. Good lift-off was accomplished via using two photoresist layers, LOR10 and AZ 5214E respectively, and conventional contact optical photolithography methods were used to define all layers. All finalized wafers were then diced, to attain 5x5mm$^2$ memristor chips, which were wire-bonded in standard packages for measurements. Preliminary characterization of all samples took place on wafer by employing a Wentworth semi-automatic prober and a Keithley SCS-4200 semiconductor characterization suite.

The cross-section of our memristor prototypes appearing on the inset of Figure 1a is a 256x256 pixel EDX map of a pristine (as-fabricated) device. This map was taken at 50μs dwell time, 1.2nA beam current and 8mins acquisition time on a FEI Titan G2 ChemiSTEM 80-200 microscope.

**Supplementary Information** is available in the online version of the paper.

**Acknowledgements** We acknowledge the financial support of the eFutures XD EFXD12003-4, the CHIST-ERA ERA-Net and EPSRC EP/J00801X/1, EP/K017829/1 and FP7-RAMP. We are also grateful to Profs. Jesper Sjöström and Guoqiang Bi for giving us permission to use their data in this manuscript and for providing detailed information on the experimental protocols used to acquire their data. Their support was essential for benchmarking our experimental set ups and data against real biological synapses.

**Author Contributions** T.P. and E.V. conceived the experiments. T.P. and A.K. fabricated the samples. E.V. modeled the plasticity phenomena. S.L.W., R.B. and I.S. performed the electrical characterization of the samples. All authors contributed in the analysis of the results and in writing the manuscript.

**Author Information** Reprints and permissions information is available at. The authors declare no competing financial interests. Correspondence and requests for materials should be addressed to T.P. (t.prodromakis@soton.ac.uk).

# Figure 1

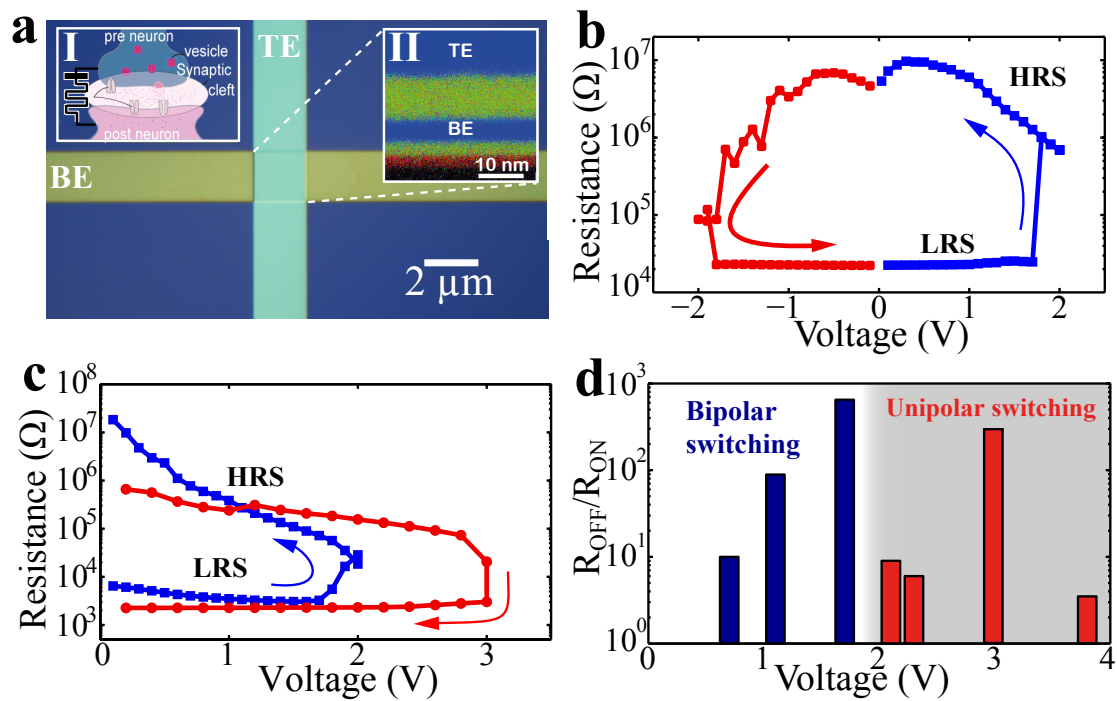

# Figure 2

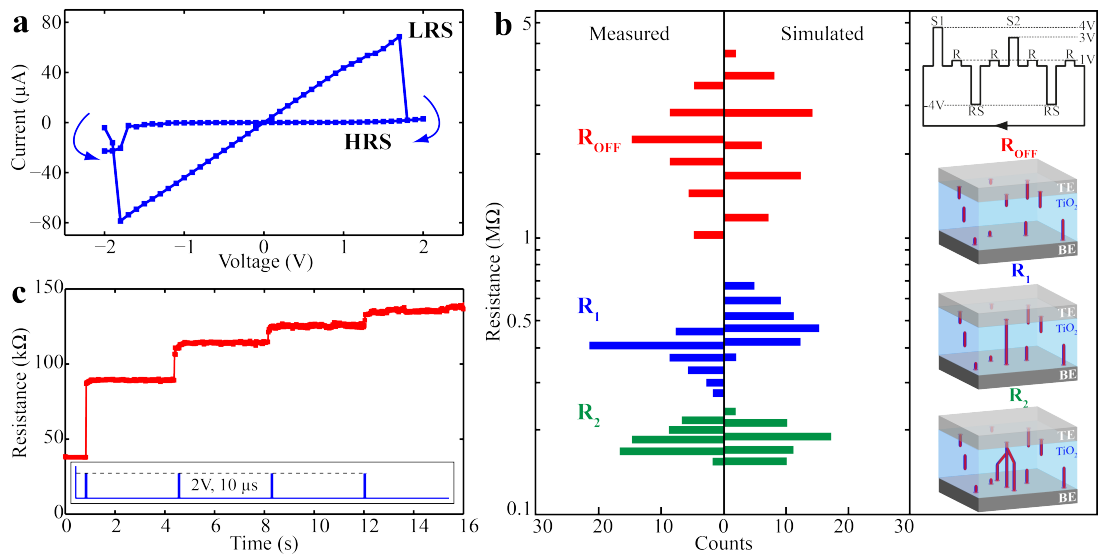

# Figure 3

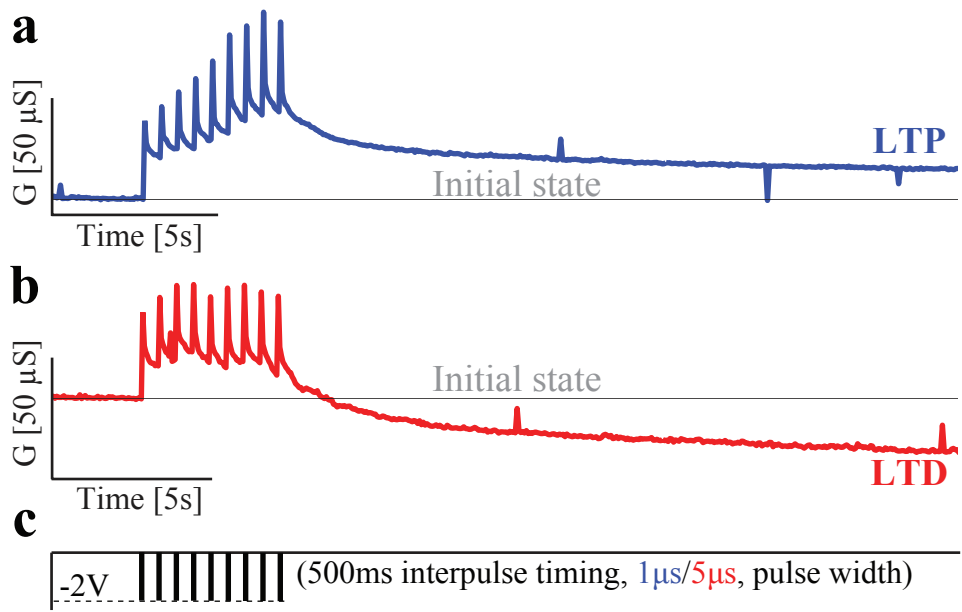

## Figure 4

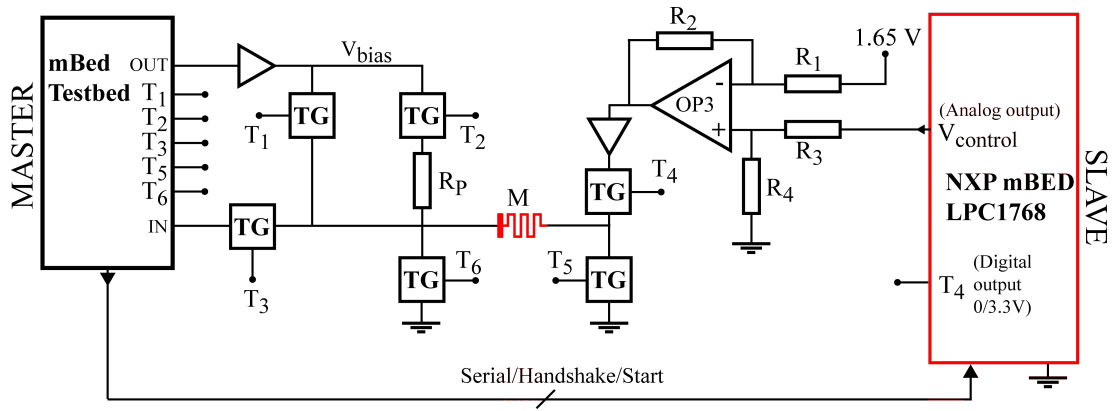

## Figure 5

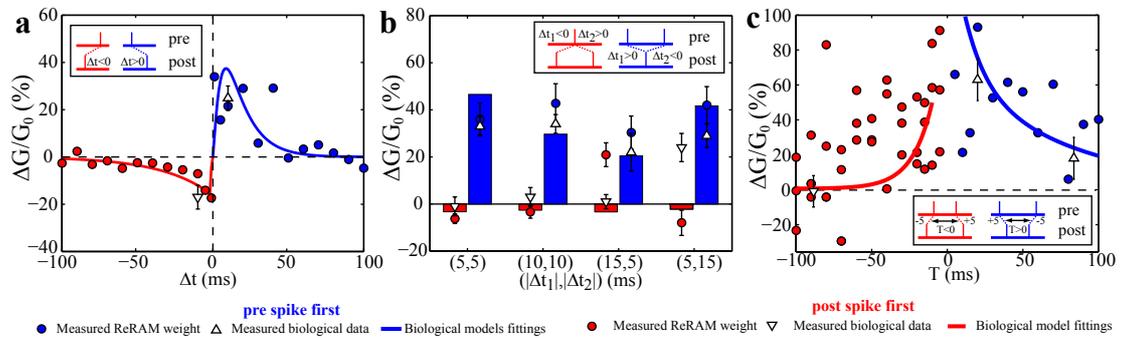

## Figure 6

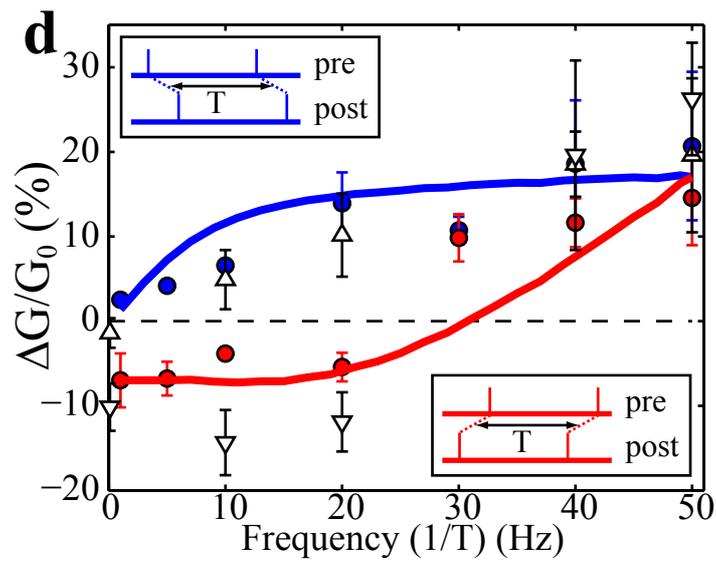